\documentclass[aps,twocolumn,amsmath,amssymb,preprintnumbers]{revtex4}
\usepackage{amsmath} \usepackage{amsfonts} \usepackage{amssymb}
\usepackage{bbm}
\usepackage{epsfig}
\usepackage{graphics}
\usepackage{graphicx}
\newcommand{\be}{\begin{equation}}
\newcommand{\ee}{\end{equation}}
\newcommand{\ba}{\begin{eqnarray}}
\newcommand{\ea}{\end{eqnarray}}
\newcommand{\nn}{\nonumber}

\newcommand{\kl}{\langle}
\newcommand{\kr}{\rangle}

\begin{document}

\title[ ]{Quantum entanglement and interference from classical statistics}

\author{C. Wetterich}
\affiliation{Institut  f\"ur Theoretische Physik\\
Universit\"at Heidelberg\\
Philosophenweg 16, D-69120 Heidelberg}

\begin{abstract}
Quantum mechanics for a four-state-system is derived from classical statistics. Entanglement, interference, the difference between identical fermions or bosons and the unitary time evolution find an interpretation within a classical statistical ensemble. Quantum systems are subsystems of larger classical statistical systems, which include the environment or the vacuum. They are characterized by incomplete statistics in the sense that the classical correlation function cannot be used for sequences of measurements in the subsystem.
\end{abstract}

\maketitle

A classical description of quantum mechanics has been tried many times, partly motivated by conceptual difficulties of understanding simple quantum phenomena, as the Einstein-Rosen-Podolski (EPR) paradox \cite{EPR}. In particular, it was hoped that a deterministic description in terms of ``hidden variables'' may be possible. In present days, however, it is common belief that the phenomenon of entanglement, which is the center-piece of the EPR-paradox, is genuinely of a quantum mechanical nature and differentiates quantum statistics from classical statistics. Entanglement is the key stone for quantum computing \cite{Zo} and the basis of theoretical considerations of the foundations of quantum mechanics \cite{Zu}. Beautiful experiments employ entanglement for teleportation \cite{Ze}.

One may take the attitude that the fundamental description of the world must be of a probabilistic nature \cite{GenStat}, with deterministic laws arising only in limiting (albeit quite genuine) cases. In this context we propose here that quantum statistics can be obtained from classical statistics. On one side it is not a deterministic ``hidden variable theory'', on the other side it does not need any concept beyond a classical statistical ensemble. The differences between classical statistics and quantum statistics are only apparent - they are due to particularities of an ``incomplete'' statistical description of subsystems \cite{3}. No additional concepts are needed beyond the classical statistical ensemble with a probability distribution for classical states, fixed values of classical observables in these states, and appropriate conditional probabilities for the outcome of sequences of measurements. 

Several attempts have shown quantum mechanical features in classical statistical ensembles \cite{3}, \cite{4}, but a full implementation of all features of quantum mechanics within classical statistics has not been achieved previously. In a sense, the functional integral representation of quantum field theory is already quite close to classical statistics \cite{ZJ}. The euclidean version after analytic continuation is often given by a standard classical statistical ensemble with a positive probability distribution. Unfortunately, this necessary positivity of the probability is lost in the formulation in Minkowski space with real time - a complex phase factor replaces the euclidean probability distribution.

If a classical statistical description of quantum  mechanics is possible, an explicit construction should be possible for the simplest systems with two or four states. Since four states can already describe an entangled state between two two-state subsystems, we concentrate in this note on this case. We explicitly present a classical statistical ensemble which realizes all laws of quantum mechanics for the four-state system, and demonstrate how entanglement and interference arise in this classical statistical setting. 

A classical statistical ensemble which describes four-state quantum mechanics typically involves infinitely many classical states $\tau$, with probabilities $p_\tau\geq 0$, $\sum_\tau p_\tau=1$. Such an ensemble describes the quantum system together with its environment or the vacuum. The quantum system can be regarded as an isolated subsystem, where the meaning of ``isolation'' will be discussed later. Only a small part of the information contained in the probability distribution $\{p_\tau\}$ is needed for a description of the subsystem, while the remaining part characterizes the environment or the vacuum. More precisely, a pure state of the subsystem can be described by a set of real numbers $f_k$, which correspond to expectation values of observables that can be measured in the subsystem, without invoking detailed knowledge of the environment. We assume that all properties of the subsystem can be described by the set $\{f_k\}$. For four-state quantum systems fifteen numbers $f_k$ (which obey additional constraints) will be needed. Obviously, the specification of fifteen expectation values is only a very small part of the information contained in the infinity of probabilities $p_\tau$. While all $f_k$ can be computed from $\{p_\tau\}$, the inverse is not possible since many different probability distributions $\{p_\tau\}$ lead to the same set $\{f_k\}$. 

\medskip\noindent
{\bf Probabilistic observables}

We concentrate on possible measurements that can only resolve two bits. (In a quantum language this corresponds to two spins that can only have the values up or down.) For any individual measurement, the measurement-device or apparatus can only take the values $+1$ or $-1$ for bit $1$ and the same for bit $2$. In total there are four possible outcomes of an individual measurement, i.e. $(++),(+-),(-+)$ and $(--)$. We describe such measurements by {\em probabilistic observables} $A$ that are characterized by the probabilities $w^{(A)}_+(f_k)$ and $w^{(A)}_-(f_k)=1-w^{(A)}_+(f_k)$ to find a value $+1$ or $-1$ in any given state of the subsystem. These probabilities only depend on $f_k$. In contrast, for any classical state $\tau$, the observable has a fixed value $A_\tau=\pm 1$, such that $A^2_\tau=1$ for all states. The probabilities $w^{(A)}_\pm$ to find values $A=\pm 1$ are related to the expectation value of $A$ in a given state of the subsystem $\{f_k\}$
\be\label{1}
\kl A\kr(f_k)=\sum_\tau p_\tau A_\tau=w^{(A)}_+(f_k)-w^{(A)}_-(f_k).
\ee
We note that $\kl A^2\kr(f_k)=w^{(A)}_+(f_k)+w^{(A)}_-(f_k)=1$, as it should be for an observable that can only take the values $\pm 1$ for any individual measurement. (See ref. \cite{N} for more details about probabilistic observables and \cite{10} for the option that probabilistic observables can be used as the fundamental statistical definition of observables.) We assume that $A$ is a ``system observable'' in the sense that $\kl A\kr$ depends only on $f_k$ and not on properties of the environment. 

We focus first on three such ``two-level observables'', namely $T_1$ for the measurement of bit $1,~T_2$ for the measurement of bit $2$, and $T_3$ for the measurement of the product of bit $1$ and bit $2$. Denoting by $w_{++},w_{+-},w_{-+}$ and $w_{--}$ the probabilities to measure in the ensemble the outcomes $(++),(+-),(-+)$ and $(--)$, one has 
\ba\label{3}
\langle T_1\rangle&=&w_{++}+w_{+-}-w_{-+}-w_{--}\nn\\
\langle T_2\rangle&=&w_{++}-w_{+-}+w_{-+}-w_{--}\nn\\
\langle T_3\rangle&=&w_{++}-w_{+-}-w_{-+}+w_{--},
\ea
such that $w_{++}$ etc. can be found from the average values of the three observables $\langle T_m\rangle$. An eigenstate of the observable $A$ has a fixed value for all measurements, or a vanishing dispersion $\langle A^2\rangle-\langle A\rangle^2=0~,~\langle A\rangle=\pm 1$. For $\langle T_1\rangle=1$ the probability $p_\tau$ for all classical states for which $A_\tau=-1$ must vanish. Such a state leads to $w_{-+}=w_{--}=0$.

\medskip\noindent
{\bf Pure and mixed states}

Let us now specify our subsystem. For the manifold of all pure states of the subsystem we choose the homogeneous space $SU(4)/SU(3)\times U(1)$. We parameterize the embedding space ${\mathbbm R}^{15}$ by the $15$ components $f_k$ of a vector $(k=1\dots 15)$. It is normalized according to $\sum_kf^2_k=3$, and obeys eight additional constraints that reduce the independent coordinates to six, as appropriate for the dimension of the complex projective space $CP^3=SU(4)/SU(3)\times U(1)$. (For a geometrical discussion of the complex projective spaces for pure states in quantum mechanics cf. ref. \cite{BH}.) An easy way to obtain the constraints for $f_k$ employs a hermitean $4\times 4$ matrix $\tilde \rho$,
\be\label{5}
\tilde\rho=\frac14(1+f_kL_k)~,~f_k=\text{tr}(\tilde\rho L_k).
\ee
(Summation over repeated indices is always implied.) Here $L_k$ are fifteen hermitean $4\times 4$ matrices obeying
\ba\label{6}
L^2_k=1~,~\text{tr} L_k=0~,~\text{tr}(L_kL_l)=4\delta_{kl}.
\ea
They read explicitly (with $\tau_k$ the Pauli $2\times 2$ matrices)
\ba\label{5A}
L_1&=&\text{diag}(1,1,-1,-1)~,~L_2=\text{diag}(1,-1,1,-1)~,\nn\\
L_3&=&\text{diag}(1,-1,-1,1)~,~L_4=\left(\begin{array}{cc}\tau_1,&0\\0,&\tau_1
\end{array}\right),\\
L_5&=&\left(\begin{array}{cc}\tau_2,&0\\0,&\tau_2
\end{array}\right),
L_6=\left(\begin{array}{cc}\tau_1,&0\\0,&-\tau_1
\end{array}\right),
L_7=\left(\begin{array}{cc}\tau_2,&0\\0,&-\tau_2
\end{array}\right),\nn
\ea
with $L_8,L_9,L_{10},L_{11}$ obtained from $(L_4,L_5,L_6,L_7)$ by exchanging the second and third rows and columns, and $L_{12},L_{13},L_{14},L_{15}$ similarly by exchange of the second and fourth rows and columns. The matrix $\tilde \rho$ parameterizes the homogeneous space $SU(4)/SU(3)\times U(1)$ if it obeys
\be\label{7}
\tilde\rho=U\hat\rho_1U^\dagger~,~UU^\dagger=U^\dagger U=1~,~\hat\rho_1=\text{diag}(1,0,0,0),
\ee
for some appropriate unitary matrix $U$. This implies
\be\label{8}
\tilde\rho^2=\tilde\rho~,~\tilde\rho^2_{\alpha\alpha}\geq 0~,~\sum\nolimits_{\alpha}\tilde \rho_{\alpha\alpha}=\text{tr}\tilde\rho=1.
\ee

The observables $T_{1,2,3}$ are specified by $\kl T_m\kr(f_k)=f_m~,~m=1,2,3$, which is equivalent to the specification of $w^{(T_m)}_\pm(f_k)$ in eq. \eqref{1}. Already at this stage we get a glance on the possibility of entanglement, since  states with $f_1=f_2=0~,~f_3=-1$ will lead to $\langle T_1\rangle=\langle T_2\rangle=0$, $\langle T_3\rangle=-1$, and therefore to a correlation for opposite values of bit $1$ and bit $2,~w_{++}=w_{--}=0~,~w_{+-}=w_{-+}=\frac12$. We label these two-level observables by a real vector with components $e_k$, with $e_k(T_m)=\delta_{km},m=1\dots 3~,~k=1\dots 15$. The mean value of $T_m$ in a given state $f_k$ can then be written in the form (with $e^{(m)}_k\equiv e_k (T_m)$)
\be\label{9}
\kl T_m\kr(f_k)=f_ke^{(m)}_k.
\ee
We next represent an observable $A(e_k)$, labeled by $e_k$, in terms of a hermitean operator
\be\label{9A}
\hat A=e_kL_k~,~e_k(A)=\frac14\text{tr}(\hat AL_k),
\ee
with $\sum_ke^2_k=1$ for $\hat A^2=1$. In this language one finds $\kl T_m\kr=\text{tr}(\hat T_m\tilde \rho)$, $\hat T_m=L_m$.  

The states labeled by $f_k$ will correspond to pure states in the four-state quantum system. Mixed quantum states obtain, for example, by the replacement $f_k\to\rho_k$ with
\be\label{10A}
\rho_k=\left(\frac{P}{3}\right)^{1/2}f_k,
\ee
and purity $P$ obeying
\be\label{10B}
P=\rho_k\rho_k\leq 3.
\ee
The state of the subsystem is now determined by the fifteen numbers $\rho_k$ and we assume for $m=1,2,3$ that $\kl T_m\kr=\rho_m$. For given $\rho_k$ the state of the subsystem can be described by a density matrix $\rho$, 
\be\label{11}
\rho=\frac14(1+\rho_kL_k).
\ee
 In terms of $\rho$ one finds the familiar quantum law for expectation values
\be\label{12}
\langle A\rangle=\text{tr}(\hat A\rho)
\ee
for every observable whose expectation value obeys $\kl A\kr=\rho_ke_k$, as for the ``one bit observables'' $T_m$. 

We also can use $\rho$ in order to formulate the general condition that the $\rho_k$ have to obey if they describe a four-state quantum system. An arbitrary hermitean $\rho$ can be diagonalized by a suitable unitary transformation
\be\label{13A}
\rho=U~diag~(p_1,\dots,p_4)U^\dagger
\ee
and we require that $\rho$ is a positive matrix,
\be\label{13B}
0\leq p_\alpha\leq 1.
\ee
The relations \eqref{13A}, \eqref{13B}, together with the definition \eqref{11}, may be called ``positivity constraint'' or ``purity constraint''. A subsystem characterized by expectation values $\rho_k$ can behave as a quantum system only if the fifteen numbers $\rho_k$ obey the purity constraint. In turn, the purity constraint ensures that the diagonal elements of the density matrix are positive and bounded by one, $0\leq\rho_{\alpha\alpha}\leq 1$. For pure states one has $P=3$ and $\rho$ coincides with $\tilde\rho$ \eqref{8}. More generally, one finds
\be\label{13C}
P=\rho_k\rho_k=4\text{tr}\rho^2-1=4\sum_\alpha p^2_\alpha-1\leq 3.
\ee
The purity is a statistical quantity characterizing the fluctuations in the subsystem. We will see that it limits the number of independent system observables which can have sharp values simultaneously, motivating the name ``purity constraint''.

The description of pure states in terms of wave functions $\psi_\alpha~,~\psi^\dagger\psi=1$, can be obtained from the density matrix in a standard way, $\tilde\rho_{\alpha\beta}=\psi_\alpha\psi^*_\beta~,~\psi_\alpha=U_{\alpha\beta}
(\psi_1)_\beta~,~(\psi_m)_\alpha=\delta_{m\alpha}$. This expresses the $f_k$ as a quadratic form in the complex four-vector $\psi_\alpha$,
\be\label{12B}
f_k=\psi^\dagger L_k\psi,
\ee
and shows directly that only six components of $f_k$ are independent. The quantum mechanical wave function $\psi$ appears here as a convenient way to parameterize the manifold of pure states. The expectation value $\kl A\kr=f_ke_k$ translates into the quantum mechanical law
\be\label{14A}
\langle A\rangle=\psi^\dagger\hat A\psi.
\ee

Assume now that the time evolution of the classical probability distribution $\{p_\tau\}$
changes a pure state of the subsystem into another pure state. It therefore corresponds to a rotation in the space of the (constrained) vectors $f_k$ which parameterize $SU(4)/SU(3)\times U(1)$. According to eq. \eqref{7} this is represented in the quantum language by a unitary evolution of the pure state density matrix
\be\label{13}
\rho(t)=U(t,t')\rho(t')U^\dagger(t,t').
\ee
The Schr\"odinger equation for the wave function can be easily derived from eq. \eqref{13}. The unitary time evolution \eqref{13} is a characteristic feature for the isolation of the subsystem. We may assume that it also holds for mixed states, i.e. all density matrices obeying the purity constraint. The consequences of this setting are simple but striking: the time evolution of expectation values of the system observables precisely follows the time evolution in quantum mechanics for an appropriate Hamiltonian $H=i\big(\partial_t U(t,t')\big)U^\dagger(t,t')$. If $H$ is independent of $\rho_k$, the quantum time evolution is linear. 

\medskip\noindent
{\bf Entangled states}

Let us next look at a typical entangled quantum state. We consider the wave functions
\be\label{14}
\psi_\pm=\frac{1}{\sqrt{2}}(\psi_2\pm\psi_3),
\ee
with associated pure state density matrices $\rho_\pm$. These states are eigenstates to $\hat T_3$ with eigenvalue $-1$. Writing
\be\label{16}
\rho_\pm=\frac14\big(1-L_3\pm(L_{12}-L_{14})\big),
\ee
we infer $f_3=-1~,~f_{12}=\pm 1~,~f_{14}=\mp 1$, and all other $f_k$ vanishing. Thus $\langle T_1\rangle=\langle T_2\rangle=0$ implies that the values of bit $1$ and bit $2$ are randomly distributed, with equal probabilities to find $+1$ or $-1$. Nevertheless due to $\langle T_3\rangle=-1$, the product of both bits has a fixed value. Whenever bit $1$ is measured to be positive, one is certain that a measurement of bit $2$ yields a negative value, and vice versa. The two bits are maximally anticorrelated. The situation is the same for ``spin observables in other directions'' that we will discuss later in this note. Obviously, an entangled state can be realized by a subsystem with the appropriate values $\rho_k=f_k$, and its time evolution will be unitary if the purity is conserved. 

At this stage we may give an explicit example for a classical statistical ensemble that realizes such an entangled state. Let us label the classical states as $\tau=\big(\{\sigma_k\},\zeta\big)$, with $\{\sigma_k\}$ an ordered sequence of fifteen discrete variables $\sigma_k=\pm 1$ which accounts for $2^{15}$ different possibilities. The number of classical states is typically higher than $2^{15}$ - this is taken into account by the additional index $\zeta$. We identify the expectation values which specify the state of subsystem as 
\be\label{16A}
\rho_k=\kl \sigma_k\kr=\sum_{\big(\{\sigma_k\},\zeta\big)}\sigma_k
p\big(\{\sigma_k\},\zeta\big),
\ee
with $\rho_k=f_k$ for pure states. In particular, the two bits or two spin variables $T_1,T_2$ are associated with the classical observables $\sigma_1$ and $\sigma_2$, while $\sigma_3$ specifies the measurement correlation of pairs of measurements for $T_1$ and $T_2$, i.e. $\kl T_2T_1\kr_m=\rho_3=\kl\sigma_3\kr$. 

The probability distributions $\{p_\tau\}$ which describe an isolated subsystem (these are not the most general allowed probability distributions) can be written as 
\ba\label{16B}
p\big(\{\sigma_k\}, \zeta\big)&=&p_s\big(\{\sigma_k\}\big)\bar p_s(\zeta)+
\delta p_e\big(\{\sigma_k\},\zeta\big),\nn\\
p_s\big(\{\sigma_k\}\big)&=&2^{-15}\prod_k(1+\sigma_k\rho_k),\nn\\
\sum_{\{\sigma_k\},\zeta}\sigma_k\delta p_e
\big(\{\sigma_k\},\zeta\big)&=&0~,~\sum_\zeta\bar p_s(\zeta)=1.
\ea
The information contained in the part of the probability distribution $\delta p_e$ describes properties of the environment, but is not needed for the computation of the state of the subsystem. We observe that the classical correlation functions as $\kl \sigma_k\sigma_l\kr$ typically depend on $\delta p_e$. They cannot be computed from the information available for the subsystem, i.e. in terms of $\rho_k$. The subsystem is described by ``incomplete statistics'' in this sense. We will argue below that sequences of measurements in the subsystem are described by a different conditional correlation function. 

Finally, the unitary time evolution characteristic for the isolated subsystem is realized by an evolution of $p_s(\rho_k)$ according to the unitary evolution of the $\rho_k$ corresponding to eq. \eqref{13}, while the evolution of $\delta p_e$ is arbitrary as long as the constraint \eqref{16B} is respected. The classical probability distribution for the entangled state \eqref{16} obtains by inserting in $p_s(\rho_k)$ the appropriate values $\rho_k=f_k$. Since an arbitrary unitary evolution can be described by an appropriate evolution of the classical probability distribution $\{p_\tau\}$, we can realize ``quantum operations'' as the ``CNOT-gate'' for the two bits by a suitable time evolution of a classical statistical ensemble. It is sufficient that after a given time interval the unitary matrix takes the form 
\be\label{19A}
U=\left(\begin{array}{llll}
1,0,0,0\\0,1,0,0\\0,0,0,1\\0,0,1,0
\end{array}\right).
\ee

\medskip\noindent
{\bf Interference}

Other interesting quantum phenomena are the superposition of states and interference. Just as entanglement, they can be described within our classical statistical ensemble. We concentrate on pure states $\rho_k=f_k$. Consider two pure quantum states that evolve in time according to
\be\label{17}
\psi_a=\frac{1}{\sqrt{2}}(\psi_1+\psi_2)e^{-i\omega_at}~,~
\psi_b=\frac{1}{\sqrt{2}}(\psi_1-\psi_2)e^{-i\omega_bt}.
\ee
The corresponding density matrices are time independent, 
\be\label{22A}
\rho_{a,b}=(1+L_1\pm L_4\pm L_6)/4. 
\ee
Both states describe an eigenstate of the first bit, $\langle T_1\rangle =1$, whereas the second bit is randomly distributed, $\langle T_2\rangle=0$.  Due to the time dependent phase, the interference can be positive or negative for the superposition of both states, $\psi=\frac{1}{\sqrt{2}}(\psi_a+\psi_b)$. For such a superposed state a quantum mechanical computation leads to a characteristic oscillation of $\langle T_2\rangle$,
\be\label{19}
\langle T_2\rangle=\psi^\dagger L_2\psi=\cos(\Delta t)~,~\Delta=\omega_a-\omega_b,
\ee
as known from the oscillation of a spin in the $z$-direction for a superposition of spin-eigenstates in the $x$-direction, with different energies for the positive and negative $S_x$ eigenvalues. A rotation of the $\rho_k=f_k$, describing the time evolution of the classical probability distribution $\{p_\tau\}$ according to eq. 
\eqref{16B}, with $f_2=f_3=\cos(\Delta t)$, reproduces the ``interference pattern'' \eqref{19}. An evolution law $\partial_tf_2=\Delta f_5~,~\partial_tf_5=-\Delta f_2~,~f_3=f_2$, $f_7=f_5~,~f_1=1~,~f_k=0$ otherwise, has indeed solutions leading to a density matrix which corresponds to the superposed state $\psi$,
\be\label{20}
\rho=\frac14\Big\{1+L_1+\cos(\Delta t)(L_2+L_3)-\sin(\Delta t)(L_5+L_7)\Big\}.
\ee

\medskip\noindent
{\bf Bosons and fermions}

Furthermore, our classical statistical ensemble can describe identical bosons or fermions. We may identify the two bits with two particles that can have spin up or down,
\be\label{21}
\psi_1=|\uparrow\uparrow\rangle~,~\psi_2=|\uparrow\downarrow\rangle~,~\psi_3=|
\downarrow\uparrow\rangle~,~\psi_4=|\downarrow\downarrow\rangle.
\ee
If the particles are identical, no distinction between bit $1$ and bit $2$ should be possible. This requires that the system must be symmetric under the exchange of the two bits, imposing restrictions on the allowed probability distributions $\{p_\tau\}$. The symmetry transformation corresponds to an exchange of the second and third rows and columns of $\tilde \rho$. On the level of the $f_k$ this amounts to a mapping $f_k\to f'_k~:~f_1\leftrightarrow f_2~,~f_4\leftrightarrow f_8~,~f_5\leftrightarrow f_9~,~f_6\leftrightarrow f_{10}~,~f_7\leftrightarrow f_{11}~,~f_{13}\leftrightarrow f_{15}$, while $f_3,f_{12}$ and $f_{14}$ remain invariant. Allowed probability distributions \eqref{16B} must obey $p_s(f'_k)=p_s(f_k)$. In particular, the allowed pure states are restricted by $f_1=f_2~,~f_4=f_8~,~f_5=f_9~,~f_6=f_{10}~,~f_7=f_{11}$, and $f_{13}=f_{15}$.

Consider the pure states $\psi_+$ and $\psi_-$ in eq. \eqref{14}. For both states the density matrix $\rho_\pm$  is compatible with the symmetry.  This does not hold for the density matrices corresponding to the states $\psi_2$ or $\psi_3$. In fact, linear superpositions of $\psi_+$ and $\psi_-$ are forbidden by the symmetry, a pure state $a\psi_-+b\psi_+$ must have $a=0$ or $b=0$. The symmetry requirement acts as a ``superselection rule'' for the allowed pure states or density matrices. For an arbitrary state vector $a\psi_-+b\psi_++c\psi_1+d\psi_4$ the symmetry of $\rho$ requires either $a=0$ or $b=c=d=0$. We observe that $\psi_-$ switches sign under the symmetry transformation as characteristic for a state consisting of two identical fermions. In contrast, the boson wave function $\psi=b\psi_++c\psi_1+d\psi_2$ is invariant under the ``particle exchange symmetry''. The time evolution for ``identical bits'' must be consistent with the exchange symmetry. It corresponds to $SU(3)$ rotations in the space of bosons, while the fermion state has a time independent density matrix.

\medskip\noindent
{\bf System observables}

So far we have only used the mutually commuting operators $L_1,L_2,L_3$. One may ask if other operators of quantum mechanics, as arbitrary $L_k$, can correspond to observables in our classical statistical setting. It is straightforward to find probabilistic two level observables $T_k$ corresponding to arbitrary $L_k$.  We require $\kl T_k\kr=\rho_k$ and associate $T_k$ with the classical observables $\sigma_k$ in eq. \eqref{16A}. Associating to every quantum operator the vector $e_k$ according to eq. \eqref{9A}, we specify $\kl T_k\kr=\rho_ke_k$, leading directly to the quantum rule \eqref{12} for $\langle T_k\rangle$. We can obviously find eigenstates for arbitrary $T_k$, as $\rho_a$ in eq. \eqref{22A}, which is an eigenstate to $T_4$ with eigenvalue $+1$ and similar for $T_6$. In fact, the operators $L_1,L_4,L_6$ mutually commute such that there exist simultaneous eigenstates for all three observables. On the other hand, $T_2$ cannot have a sharp value in this state - $L_2$ does not commute with $L_4$ and $L_6$. The quantum mechanical uncertainty principle for non-commuting operators is directly implemented in our classical statistical setting. If we associate $(L_1,L_2)$ to spins in the $z$-direction, it is straightforward to associate $(L_8,L_4)$ and $(L_9,L_5)$ to spins in the $x$- and $y$-directions, respectively. We mayconsider the fifteen observables $T_k$ as basis observables. A sharp value for a basis observable requires $\rho_k=\pm 1$, and we see how the purity constraint \eqref{13C} limits the number of simultaneously sharp basis observables to at most three.

While the observables $T_k$ are again two-level observables with spectrum $+1$ and $-1$ (corresponding to $L^2_k=1$), the possible classical observables associated to arbitrary hermitean operators can have a spectrum of possible values for individual measurements with up to four different real values. Furthermore, we may also wish to consider rotated two level observables in arbitrary directions. The realization of such a larger set of observables of the subsystem as classical observables with fixed values in every state $\tau$ of the classical ensemble requires infinitely many classical states $\tau$ \cite{N} (using the index $\zeta$ in eq. \eqref{16A}). One finds that arbitrary observables of the four-state quantum system can find a classical realization in this way. To all such system observables one can associate a vector $e_k$ such that $\kl A(e_k)\kr=\rho_ke_k$. Therefore the ``quantum rule'' for expectation values, $\kl A(e_k)\kr=\text{tr}(\hat A(e_k)\rho)$, with $\hat A(e_k)=e_kL_k$, holds for all system observables. In particular, the probabilities to find for arbitrary two-level observables $A(e_k)$ (represented by operators obeying $\hat A^2(e_k)=1)$ the values $+1$ or $-1$ obey $w_\pm(\rho_k;e_k)=\frac12(1\pm \rho_k e_k)$. 

One may wonder why the ``system observables'', whose expectation values obey the simple rule $\kl A\kr=\rho_ke_k$, play such an important role for the description of subsystems which show a quantum character. We invoke here the ``principle of equivalent state and observable transformations'' (PESOT). This states that it should be possible to undo any time evolution of the state by a corresponding time evolution of the observables. In the quantum mechanical language, this amounts to the well known equivalence between the Schr\"odinger and Heisenberg pictures - the time evolution can be either described by the evolution of the state or the evolution of the operators. According to PESOT one can always define an appropriate observable transformation such that $\kl A\kr$ remains invariant under a change of $\rho_k$ \cite{GenStat}. Since the $\rho_k$ transform under $SU(4)$ as the adjoint representation (with additional nonlinear constraints \eqref{7}, \eqref{10A}, \eqref{10B}), one needs another adjoint representation characterizing the transformation of the observables, namely $e_k$, in order to construct the invariant scalar product. Together with linearity in $e_k$ (or $\rho_k$) this fixes $\kl A\kr=e_k\rho_k$.

\medskip\noindent
{\bf Conditional correlations}

Beyond a rule for the computation of expectation values a probabilistic description of Nature has to specify a rule how the probability for sequences of events or sequences of measurements is computed. For a sequence of measurements of first $B$ and then $A$ (which may be associated to a sequence of two events) we need the {\em conditional probability} to find one of the possible measurement values for $A$ under the condition that a certain value of $B$ has been measured. In general, conditional probabilities depend on the precise way how measurements are performed, since the measurement of $B$ changes the subsystem and this change has to be specified. For a very large number of effective degrees of freedom, as relevant for classical thermodynamics, the influence of the measurement of $B$ on the system is often neglected. This is not valid for our description of subsystems. 

We denote the conditional probability to find a value $\epsilon$ for bit $2$ if bit $1$ has been measured to have a value $\gamma$ by $p(\epsilon;\gamma)$. For the entangled state \eqref{14}, \eqref{16} it obeys $p(1;1)=p(-1;-1)=0~,~p(1;-1)=p(-1;1)=1$. In our classical statistical setting we describe the product of two measurements of observables $A$ and $B$ by a ``measurement correlation'' $\kl AB\kr_m$. For the two bits $T_1$ and $T_2$ it is given by $(\kl T_1\kr=w_{1+}-w_{1-})$
\ba\label{16C} 
\kl T_2T_1\kr_m&=&p(1;1)w_{1+}+p(-1;-1)w_{1-}\nn\\
&&-p(1;-1)w_{1-}-p(-1;1)w_{1+},
\ea
and we can identify the probabilities in eq. \eqref{3} as $w_{++}=p(1;1)w_{1+}~,~w_{+-}=p(1;-1)w_{1-}$ etc.. In turn, for $w_{1\pm}\neq 0$ the conditional probabilities $p(\epsilon;\gamma)$ can be computed from $\kl T_2T_1\kr_m=\rho_3$ and $\kl T_1\kr=\rho_1~,~\kl T_2\kr=\rho_2$, using $p(1;\gamma)+p(-1;\gamma)=1$. Here we assume that for the special case of two ``comeasurable'' bits the expectation value $\kl T_2\kr$ is not influenced by the measurement of $T_1$, such that $\kl T_2\kr=p(1;1)w_{1+}+p(1;-1)w_{1-}-p(-1;1)w_{1+}-p(-1;-1)w_{1-}$. This will not hold for general pairs of observables, in particular not for pairs which will be represented by non-commuting quantum operators.

In general, we require that the outcome of a sequence of measurements which respect the isolation of the subsystem should be predictable for a given state of the subsystem, and not involve details of the environment. In other words, the measurement correlation $\kl AB\kr_m$ for a pair of two ``system observables'' should be computable in terms of $\rho_k$ and be independent of $\delta p_e$. Furthermore, we require that the repetition of a measurement yields an identical result, and that good measurements keep the maximum amount of information about the subsystem which is compatible with the first two requirements. One can then show \cite{N} that the measurement correlation equals the quantum correlation, as expressed by the expectation values for the anticommutators of operators,
\be\label{16a}
\langle AB\rangle_m=\frac12\textup{tr}\big(\{\hat A,\hat B\}\rho\big).
\ee 
The probabilities for the outcome of two measurements follow therefore exactly the quantum mechanical laws and can violate Bell's inequalities \cite{11}. 

In contrast, Bells inequalities apply to classical correlation functions for classical observables \cite{12}. However, as we have seen before, the classical correlations depend on $\delta p_e$. They cannot be used for measurements which maintain the isolation of the subsystem and only measure properties of the subsystem. The classical correlations are appropriate only for measurements of properties of the subsystem together with properties of the environment, in which we are not interested. Concerning locality and realism, our classical statistical setting shares the same properties as other classical statistical systems. Correlations can be non-local - it is sufficient that they have been prepared in the past by causal processes. Correlated systems have to be considered as a whole - parts of such a system cannot be associated to independent entities, even if signals cannot be exchanged any longer between the parts. Correlations are part of reality on the same footing as expectation values of observables. The reason for the violation of Bell's inequalities by the measurement correlations can be rooted in the incompleteness of the statistical description of the subsystem even for ensembles where the description of the subsystem together with its environment is complete. 

In conclusion, we have derived four-state quantum mechanics from classical statistical mechanics. Every quantum system with arbitrary Hamiltonian and observables finds an equivalent description in terms of a classical statistical ensemble with an appropriate time evolution of the probability distribution. In particular, we have presented an explicit description of entanglement and interference, as well as fermions and bosons, within a classical ensemble. 

Our approach is not a simple rewriting of the quantum mechanical laws. Quantum statistics appears as a particular case within a much larger class of probabilistic theories for subsystems. The restriction to a manifold of pure states $SU(M)/SU(M-1)\times U(1)$, the unitary time evolution of mixed states, or the principle of equivalent state and observable transformations could be abandoned, leading to statistical models different from quantum mechanics. For example, a decrease of the purity describes decoherence \cite{13}, while in the opposite case an increase of purity results in the approach towards a pure state or ``syncoherence'' \cite{N}. Non-linear versions of quantum mechanics could be described if the unitary matrix $U(t,t')$ in eq. \eqref{13} depends on $\rho_k$. Our approach therefore provides a framework to test quantum mechanics against ``neighboring'' statistical ensembles where some of the properties leading to quantum mechanics are not obeyed exactly. Besides the conceptual and perhaps philosophical impact of a classical statistical description of quantum mechanics, it remains to be seen if our findings can also be of practical use, for example by classical computations of quantum processes or steps in quantum computing.



\begin{thebibliography}{100}
\bibitem{EPR}A. Einstein, B. Podolski, N. Rosen, Phys. Rev. {\bf 47} (1935) 777
\bibitem{Zo}R. Feynman, Int. J. Theor. Phys. {\bf 21} (1982) 467;\\
D. Deutsch, Proc. R. Soc. London {\bf A400} (1985) 97;\\
J. I. Cirac, P. Zoller, Phys. Rev. Lett. {\bf 74} (1995) 4091
\bibitem{Zu}W. Zurek, arXiv: 0707.2832
\bibitem{Ze}D. Bouwmeester, J. W. Pan, K. Mattle, M. Eibl, H. Weinfurter, A. Zeilinger, Nature {\bf 390} (1997) 575
\bibitem{GenStat}C. Wetterich, Nucl. Phys. {\bf B314} (1989) 40; Nucl. Phys. 
{\bf B397} (1993) 299
\bibitem{3}C. Wetterich, in ``Decoherence and Entropy in Complex Systems'', ed. T. Elze, p. 180, Springer Verlag 2004, arXiv: quant-ph/0212031
\bibitem{4}C. Wetterich, Phys. Lett. {\bf B399}(1997) 123
\bibitem{ZJ}J. Zinn-Justin, Quantum field theory and Critical Phenomena, Oxford University Press 1996
\bibitem{N}C. Wetterich, arXiv: 0810.0985;\\
C. Wetterich, J. Phys. 174 (2009) 012008
\bibitem{10}A. S. Holevo, ``Probabilistic and Statistical Aspects of Quantum Theory'' (Amsterdam, North Holland) 1982; WS. T.  Ali, E. Prugovecki, J. Math. Phys. {\bf 18} (1977) 219; \\
M. Singer, W. Stulpe,  J. Math. Phys. {\bf 33} (1992) 131\\
E. Beltrametti, S. Bugajski,  J. Phys. A: Math. Gen. {\bf 28} (1995) 3329; Int.~J.~Theor.~Phys. {\bf 34} (1995) 1221; \\
S. Bugajski, Int.~J.~Theor.~Phys. {\bf 35} (1996) 2229
\bibitem{BH}D. C. Brody, L. P. Hughston, J. Geom. Phys. {\bf 38} (2001) 19
\bibitem{11}J. Bell, Physica 1 (1964) 195
\bibitem{12}J. Clauser, M. Horne, A. Shimony, R. Holt, Phys. Rev. Lett. {\bf 23} (1969) 880;\\
J. Bell, ``Foundations of Quantum Mechanics'', ed. B. d'Espagnat (New York: Academic, (1971) p. 171;\\
J. Clauser, M. Horne, Phys. Rev. {\bf D10} (1974) 526;\\
J. Clauser, A. Shimony, Rep. Prog. Phys. {\bf 41} (1978) 1881
\bibitem{13}H. D. Zeh, Found. Phys. {\bf 1} (1970) 69;\\
E. Joos, H. D. Zeh, Z. Phys. {\bf B59} (1985) 273;\\
E. Joos, H. D. Zeh, C. Kiefer, D. Giulini, J. Kupsch, I.-O. Stamatescu,
``Decoherence and the appearance of the classical world'', Springer 2003; \\
W. Zurek, Rev. Mod. Phys. {\bf 75} (2003) 715
\end{thebibliography}
\end{document}